# Context Aware Multisensor Image Fusion for Military Sensor Networks using Multi-Agent system

Ashok V. Sutagundar*, S. S. Manvi**

*Department of Electronics and Communication Engineering, Basaveshwar Engineering College, Bagalkot-587102, INDIA.
**Wireless Information Systems Research Lab, Department of Electronics and Communication Engineering, REVA , Institute of Technology and Management, Yelahanka, Bangalore
ashok_ec@yahoo.com, agentsun2002@yahoo.com.

## Abstract

*This paper proposes a Context Aware Agent based Military Sensor Network (CAMSN) to form an improved infrastructure for multi-sensor image fusion. It considers contexts driven by a node and sink. The contexts such as general and critical object detection are node driven where as sensing time (such as day or night) is sink driven. The agencies used in the scheme are categorized as node and sink agency. Each agency employs a set of static and mobile agents to perform dedicated tasks. Node agency performs context sensing and context interpretation based on the sensed image and sensing time. Node agency comprises of node manager agent, context agent and node blackboard (NBB). Context agent gathers the context from the target and updates the NBB, Node manager agent interprets the context and passes the context information to sink node by using flooding mechanism. Sink agency mainly comprises of sink manager agent, fusing agent, and sink black board. A context at the sensor node triggers the fusion process at the sink. Based on the context, sink manager agent triggers the fusing agent. Fusing agent roams around the network, visits active sensor node, fuses the relevant images and sends the fused image to sink. The fusing agent uses wavelet transform for fusion. The scheme is simulated for testing its operation effectiveness in terms of fusion time, mean square error, throughput, dropping rate, bandwidth requirement, node battery usage and agent overhead.*

## 1.Introduction

Recent advancement in wireless communications and sensor technology has enabled the development of low-cost Wireless Sensor Networks (WSNs). The sensor networks can be used for various application areas such as health, military, environmental monitoring, home, etc,. A typical WSN consists of a large number of low-cost and low-energy sensors, which are scattered in an area of interest to collect observations and pre-process the observations. Each sensor node has its own communication capability to communicate with other sensor nodes or the central node (fusion center) via a wireless channel. More recently, the production of cheap CMOS cameras and microphones, which can acquire rich media content from the environment, created a new wave into the evolution of wireless sensor networks. For instance, the Cyclops imaging module is a light-weight imaging module which can be adapted to MICA21 or MICAz sensor nodes. Thus, a new class of WSNs came to the scene that can be applied to military applications and are known as Wireless Military Sensor Networks (WMSNs).

Lot of papers have been published in this research area [1]. Most of this research has, however, focused on wireless networks of sensor nodes that collect scalar data such as temperature, pressure, and humidity sensors. Such sensors generate a limited amount of information, which can be insufficient for many applications, even if a large number of sensors are deployed. As VSNs offer new opportunities for many promising applications compared to scalar sensor networks, they also raise new challenges that are not fully addressed by current research on WSNs. Camera sensors generate a huge amount of data compared to scalar sensors. Processing and transmitting such data by generally low-power sensor nodes is challenging due to their computational and bandwidth requirements. It is emphasized in [2] that these applications demand a re-consideration of the computation communication paradigm of traditional WSNs, which has mainly focused on reducing the energy consumption, targeting to prolong the longevity of the sensor network. Though, the applications implemented by WMSNs have a second goal, as important as the energy consumption, to be pursued. This goal is the delivery of application-level quality of service (QoS) and the mapping of this requirement to network layer metrics, like latency. A VSN with overlapped field of views could exploit the redundancy between the field of view of each visual sensor to avoid inconsistencies and obtain more accurate results. Therefore, a key challenge in visual sensor network contexts is how to get the most relevant information from the environment and fuse it in the most efficient way. However getting the most relevant information from each visual sensor is not a simple task. Multiple factors could affect the visual sensor information, for example in tracking activities, occlusions of static objects could affect the tracking positions.

Context awareness can then be defined as detecting a user's internal or external state. Context-aware computing describes the situation of a wearable or mobile computer being aware of the user's state, surroundings, and modifying its behavior based on this information [3]. Context awareness plays a significant role in MSNs because it allows for interpreting various contexts such as temporal, emergency and computational contexts coming from the battlefield based on information regarding the current state of the object movement and the state of the environment. Context-aware sensing is an integral part of the MSNs design to achieve the ultimate goal of long-term pervasive enemy battle field monitoring.

This paper proposes a Context Aware Agent based Military Sensor Network (CAMSN) to form an improved infrastructure for multi-sensor image fusion. It considers contexts driven by a node and sink. The contexts such as general and critical object detection are node driven where as sensing time (such as day or night) is sink driven. The agencies used in the scheme are categorized as node and sink agency. Each agency employs a set of static and mobile agents to perform dedicated tasks. Node agency performs context sensing and context interpretation based on the sensed image and sensing time. Node agency comprises of node manager agent, context agent and node blackboard (NBB). Context agent gathers the context from the target and updates the NBB, Node manager agent interprets the context and passes the context information to sink node by using flooding mechanism. Sink agency mainly comprises of sink manager agent, fusing agent, and sink black board. A context at the sensor node triggers the fusion process at the sink. Based on the context, sink manager agent triggers the fusing agent. Fusing agent roams around the network, visits active sensor node, fuses the relevant images and sends the fused image to sink. The fusing agent uses wavelet transform for fusion.

## 1.1 Related works

Some of the related works are as follows. The focus of article [4] is on the military requirements for flexible wireless sensor networks. Based on the main networking characteristics and military use-cases, insights into specific military requirements for flexible wireless sensor networks are discussed. The article structures the evolution of military sensor networking devices by identifying three generations of sensors along with their capabilities. Existing developer solutions are presented and an overview of some existing tailored products for the military environment is given. The work presented in [5] investigates the design tradeoffs for using WSN for implementing a system, which is capable of detecting and tracking military targets such as tanks and vehicles. Such a system has the potential to reduce the casualties incurred in surveillance of hostile environments. The system estimates and tracks the target based on the spatial differences of the target object signal strength detected by the sensors at different locations. The work depicted in paper [6] proposes to use the mobile agent paradigm for reducing and aggregating data in planar sensor network architecture. Mobile agents can be used to greatly reduce the communication cost, especially over low bandwidth links, by moving the processing function to the data rather than bringing the data to a central processor.

The work given in paper [7] describes the use of the mobile agent paradigm to design an improved infrastructure for data integration in distributed sensor network (DSN). Instead of moving data to processing elements for data integration, as is typical of a client/server paradigm, it moves the processing code to the data locations. This saves network bandwidth and provides an effective means for overcoming network latency, since large data transfers are avoided. The work given in [8], describes the image transmission problem in sensor networks. It pre-processes the images in the sensors before sending them to the server, but this preprocessing requires extra energy in the sensors. In [9], a hierarchical multi-quality image fusion method based on region-mapping is proposed to improve quality of the image. Camera view is divided into regions according to mapping relation and the structured deployment of nodes. The methods of motion attention analysis and nearest neighbor sampling are also adopted to optimize the local attention region of the image. The fusion method is analyzed    and verified to improve the quality of the monitor image with no need for more energy and bandwidth. The work presented in [10] gives architecture to implement scene understanding algorithms in the visual surveillance domain. The main objective is to obtain a high level description of the events observed by multiple cameras not to fuse the tracking information. In [11], authors present a visual sensor network system with overlapped field of views, modeled as a network of software agents. The communication of each software agent allows the use of feedback information in the visual sensors, called active fusion. The work limits only to indoor scenario only.

### 1.2 Our Contributions

However most of the related works focus on how to solve different visual sensor problems, there are few works that focus on building a software architecture which allows a context aware wireless visual sensor network, where agents and wavelet transform can be applied to take active part in the fusion process. Our contributions in this paper are as follows: 1) contexts are gathered from the target (sensor node) as well as generated at the sink,  2) several static agents are defined for gathering and interpretation of the context,  3) militant activities can be monitored using the proposed MSNs, 4) used mobile agents to perform several tasks that aid information processing, fusing, etc., in asynchronous fashion, 5) based on the context, low/high resolution image fusion is employed by changing agent code, 6) embedding wavelet code in the agents for fusing images, 7) fuses only images from active sensor nodes,  and 8) reduces communication overheads and energy.

The rest of the paper is organized as follows. In section 2, we present context aware computing. Section 3 depicts the wavelet based multi-sensor image fusion. Section 4 describes context aware based information fusion. Section 5 presents the simulation model and performance metrics. Section 6 explains results and section 7 concludes the work.

## 2. Context Aware Computing

Context is any information that can be used to characterize the situation of an entity. An entity is a person, place, or object that is considered relevant to the interaction between a user and an application, including the user and applications themselves. In other words, any information that depicts the situation of a user can be entitled context. This would include the number of people in the area, the time of day, and any devices the user may employ. One can however distinguish between those contextual characteristics that are critical to the application versus those that may be relevant but are not critical. Within computing applications, there are three major context categories of interest: user context, computing resources, and environmental aspects. Orthogonal to this view, context can be explicit (that is, information provided directly by the user) or implicit (derived from on one hand sensors, on the other hand from an analysis of user behavior).

In previous decades, a narrow aspect of user context, the user preferences in terms of search and retrieval of data was denoted in the database community with the concept of views. Today, user context is considered far more broadly and includes user interests, goals, social situation, prior knowledge and history of interaction with the system. Thus, the dimension of time may also be included in the user's context. ``Context information may be utilized in an immediate, just in time, way or may be processed from a historical perspective'' [12, 13]. Environmental contextual aspects include (but are not limited to) location, time, temperature, lighting conditions, and other persons present. Note that there is some ambiguity in the literature in the use of the term environment, which can refer to the computing environment as well as the actual physical environment [14,15,16,17] of the user. Context information is often acquired from unconventional heterogeneous sources, such as motion detectors or GPS receivers. Such sources are likely to be distributed and inhomogeneous. The information from these sensors must often be abstracted in order to be used by an application; for example, GPS data may need to be converted to street addresses. Finally, environmental context information must be detected in real time and applications must adapt to changes dynamically.

To monitor the environment for changes and to adapt the behavior of the node to the current environment CAC can be used. The applications and protocols need not be aware of the environment at all, but rather focus on taking care of the tasks they have been designed for in the first place. As the network environment changes, the node must be able to rapidly adapt to the new situation. In this work, context image sensor nodes are configured based on the context. In order to reduce the communication overhead, we consider the wavelet based image fusion in WSNs.

## 3.Wavelet Based Multi Sensor Image Fusion

In this section, we describe the wavelet transform and wavelet based image fusion.

### 3.1Wavelet Transform

Wavelets are mathematical functions that cut up data into different frequency components, and then study each component with a resolution matched to its scale. They have advantages over traditional Fourier methods in analyzing physical situations where the signal contains discontinuities and sharp spikes. Wavelets were developed independently in the fields of mathematics, quantum physics, electrical engineering, and seismic geology. Interchanges between these fields during the last decade has led to many new wavelet applications such as image compression, turbulence, human vision, radar, and earthquake prediction. The wavelet transform has become a useful computational tool for a variety of signal and image processing applications. For example, the wavelet transform is useful for the compression of digital image files, smaller files are important for storing images using less memory and for transmitting images faster and more reliably. Wavelets are functions that satisfy certain mathematical requirements and are used in representing data or other functions. Wavelet algorithms process data at different scales or resolutions [20, 21, 22, 23].

### 3.2Image Fusion

Image fusion is the processing of images about a given region obtained from different sensors by a specific algorithm, so that the resultant image is more reliable, clear and more intelligible. Image fusion can take place on pixel-level, feature-level, and decision-level. Pixel-level image fusion is the basic for other levels and multiresolution image fusion based on multi-scale decomposition is a main research branch in it.

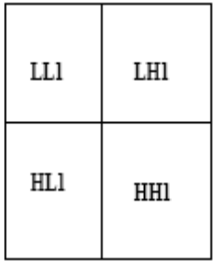 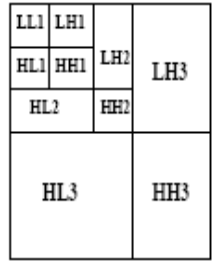 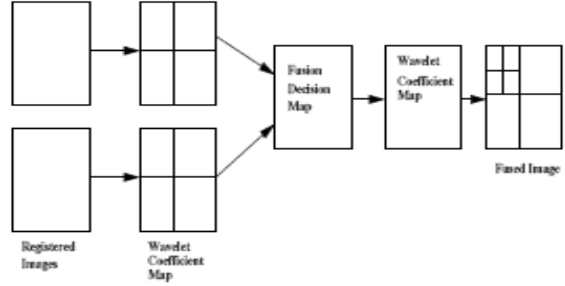

Figure 1: Sub band coding,                    Figure 2: Image Fusion using DWT

The 1-D wavelet transform can be extended to a two-dimensional (2-D) wavelet transform using separable wavelet filters. With separable filters the 2-D transform can be computed by applying a 1-D transform to all the rows of the input, and then repeating on all of the columns. The wavelet transforms of the input images are appropriately combined, and the new image is obtained by taking the inverse wavelet transform of the fused wavelet coefficients. An area-based maximum selection rule and a consistency verification step are used for feature selection. The original image of a one-level (K=1), 2-D wavelet transform, with corresponding notation is shown in figure 1.

The 2-D subband decomposition is just an extension of 1-D subband decomposition. The entire process is carried out by executing 1-D subband decomposition twice, first in one direction (horizontal), then in the orthogonal (vertical) direction. For example, the low-pass subbands (Li) resulting from the horizontal direction is further decomposed in the vertical direction, leading to LLi and LHi subbands. To obtain a two-dimensional wavelet transform, the one-dimensional transform is applied first along the rows and then along the columns to produce four subbands: low-resolution, horizontal, vertical, and diagonal. (The vertical subband is created by applying a horizontal high-pass, which yields vertical edges.) At each level, the wavelet transform can be reapplied to the low-resolution subband to further decorrelate the image. Figure 3 depicts the image fusion using DWT. DWT is first performed on each source images, then a fusion decision map is generated based on a set of fusion rules. The fused wavelet coefficient map can be constructed from the wavelet coefficients of the source images according to the fusion decision map. Finally the fused image is obtained by performing the inverse wavelet transform. When constructing each wavelet coefficient for the fused image, we have to determine which source image describes this coefficient better. This information will be kept in the fusion decision map. The fusion decision map has the same size as the original image. Each value is the index of the source image which may be more informative on the corresponding wavelet coefficient and thus, make decision on each coefficient.

Assume that node i and node j have the common information to be sent to the sink node. In order eliminate the common information between the neighboring nodes, a fusion process is adopted. Multi sensor image fusion using wavelet transform is described as follows (1) and (2).

$$F_i(u, v) = DWT(f_i(x, y), \qquad F_j(u, v) = DWT\left(F_j(x, y)\right) \qquad (1)$$

$$F_{ij}(u, v) = fusion rules\left(F_i(u, v)\right).\left(F_j(u, v)\right) \qquad (2)$$

The fused image is given by (3)

$$f_{ij}(x, y) = idwt\left(F_{ij}(u, v)\right) \qquad (3)$$

where, $f_i(x,y)$: Image of sensor node i, $f_j(x,y)$: Image of sensor node j, $f_{ij}(x,y)$: Fused Image of sensor node i and sensor node j, $F_i(u,v)$: Node i image in transform domain, and $F_j(u,v)$: Node j image in transform domain.

# 4.Context Aware Based Image Fusion

The scheme comprises of three phases: context gathering, context interpretation, and image fusion. Context gathering - contexts are gathered from the target, i.e., sensed image and time from the target, and stored in the node for a short period until its interpretation is done. Context interpretation sensed images are compared with previous image and set of critical image features (weapons, explosives, enemy, etc.) stored at the node. If image analysis yields some general or critical object feature existence, information fusion process is invoked. Information fusion- relevant images from active sensor nodes corresponding to object existence are fused to get a clear picture of the object and make some decisions.

Image fusion can classify into two types namely low resolution image fusion and high resolution image fusion. Low resolution image fusion is used for contexts like general object detection, and general image gathering by sink. High resolution image fusion is done for contexts such as critical object detection, image gathering in night time by sink. During nights, it is better to monitor the target periodically by the deployed MSN since lighting condition is poor and possibility of enemy attack, militant activities, etc,. are high. Sink driven image fusion is based on the time of sensing, available network bandwidth and sensor node battery.

Static and mobile agents are employed to perform the fusion process. The scheme assumes that an agent platform is available in the nodes of WSNs. However, if an agent platform is unavailable, the agent communicates by traditional message exchange mechanisms such as message passing method. Several agencies exist at each of the nodes based on their role in MSN, which will be discussed in this section. Following are the assumptions made in proposed work.

1. Context Interpretation is based on the image detection. Image detection is not in the scope of this work.
2. Based on the lighting condition, camera of the sensor node automatically configures itself to get the better quality images.
3. Flooding protocol is used to pass the context information to the sink node [25].
4. Geographical routing protocol is used by mobile agent to reach the active sensor nodes [26].
5. Low and high resolution image fusion code is used to fuse the images.

This section presents system environment and the agencies at the sensor and sinks nodes.

## 1.1 System Environment

Sensor node comprises of static sensor nodes and sink nodes (also called as end nodes that require information). A sensor node is said to form a cluster around it based on the communication range. Each cluster will have cluster head node. Figure 3 depicts the WSN environment. A sensor node may have several channels with different sensory devices connected to each of them. Sensor nodes are geographically distributed and periodically collect measurements of different modalities such as acoustic, seismic, and infrared from the environment. Every sensor node of WSN has predetermined value of the signal strength. Once the information is sensed by sensor node, it compares the information signal strength with predetermined value, if it is greater (if the deviation is more), a message is sent to its cluster head saying that it is an active node.

The signal energy from each channel can be detected individually and processed in the analog front end. The amount of signal energy that reaches an individual sensor is an effective indicator of how close the node is to a potential target. Once the signal is captured and pre-processed by a sensor node, the strength level of the detected signal is broadcast (through an omni directional antenna) to its cluster nodes if it greater than some predefined value.

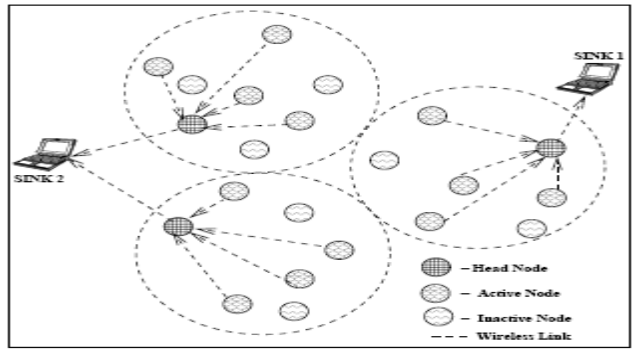

Figure 3: Distributed Sensor Network with Three Clusters

## 1.2 Agent Technology

The traditional programming paradigm uses functions, procedures, structures and objects to develop software for performing a given task. This paradigm does not support development of flexible, intelligent and adaptable software's, and also does not facilitate all the requirements of Component Based Software Engineering (CBSE) [27,28]. In recent developments, agent technology is making its way as a new paradigm in the areas of artificial intelligence and computing, which facilitates sophisticated software development with features like flexibility, scalability and CBSE requirements [29,30,31].

Agents are the autonomous programs situated within an environment, which sense the environment and act upon it to achieve the goals. Agents have special properties such as mandatory and orthogonal properties. The orthogonal properties provide strong notion of the agents [32, 33, 34]. The mandatory properties are as follows: autonomy, decision-making, temporal continuity, goal oriented. The orthogonal properties are as follows: mobility, collaborative, learning. A mobile agent platform comprises of agents, agent server, interpreter and transport mechanisms.  We classify the agent technologies as single-agent systems and muti-agent systems. In the context of single-agent systems *Local or user interface agents and Networked agents* can be identified, while in the area of multi-agent systems  *DAI (Distributed Artificial Intelligence)-based agents and Mobile agents* can be distinguished. In single-agent systems, an agent performs a task on behalf of a user or some process, while performing a task; the agent may communicate with the user as well with local or remote system resources. In contrast, the agents in multi-agent systems (MAS) may extensively cooperate with each other to achieve their individual goals, and also may interact with users and system resources.

## 1.3 Image Signal Strength

Image signal strength is measured by estimating the difference between previous image and present image stored at the sensor node. To measure the change in the image, entropy of the difference image between the present and previous image is taken. The notion of entropy may be used to estimate the information content. For an image, the simplest idea is to create these states that correspond to the possible values in which all pixels are involved. The image entropy would be given by [35]:

$$H = -\sum_{k=0}^{255} (P_k \log_2 P_k) \qquad (4)$$

Where $P_k$ is the probability of gray level k, k=0,1,2,3,…,255, assuming an 8-bit image. Equation (4) represents the information content of an image. If a change in an image takes place, the entropy, H, changes as well. Otherwise no change is detected. Firstly, subtract the previous image from a sensed image to obtain the features which is given by (5).

$$\Delta I_j = I_j - I_0 \qquad\qquad (5)$$

Where $I_j$ is the present image, $I_0$ is the previous image of the node. Entropy is then applied to measure the change in information. This step provides the image signal strength estimation, which is expressed by equation (6).

$$P_j = H(\Delta I_j) \qquad\qquad (6)$$

Where $P_j$ is difference image and H denotes entropy. Only the gray scale information of the images is used to calculate the image entropy. A threshold signal of the image is considered to find out the required signal strength where it is defined as the amount of information change from present to the previous image. This facilitates to decide the active mode of a sensor node. If $P_j$ is greater than threshold, then that node is considered to be active otherwise inactive.

### 1.4 Fused Image Transmission Model

Let $R_1, R_2, R_3,……… R_k$ be the sensor nodes deployed in the field, and $\rho$ is Fusion Factor, and $N_1$ is the Fused Image at $R_1$ and later fuses with image from other nodes, i.e., shown in equation (7).

$$N1 = Image(R1);$$

$$N1 = N1 + \sum_{k=2}^{n} \rho R_k \qquad\qquad (7)$$

Figure 4 shows the network path representation. We assume that most of the sensor nodes carry the similar images therefore instead of sending the individual images from each of the sensor nodes, the images are fused in order to remove the redundancy and fused image $f_{ij}(x,y)$ is being sent to the sink node.

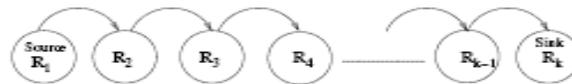

Figure 4: Network Path Representation

Agent carries the fused images from node to node depending on its given itinerary or chosen itinerary. Total transmission load, $T_{Load}$ can be given by (8).

$$TLoad = m_{al} \times hc \qquad\qquad (8)$$

where, **hc:** Number of hops required to transmit the fused image to sink node, **$m_{al}$**: Total number of packets required to transmit the image and fusion code and $T_{Load}$: Transmission Load.

## 1.5 Agencies

Each node in MSN comprises of an agent platform and the proposed agent information fusion model. The sensor nodes and sink node comprises of the node agency and sink agency respectively. Here we describe both the agencies.

### 1.5.1 Node Agency

Node agency comprises of static agents and a node blackboard (NBB) for inter agent communication. Agents are Node Manager Agent (NMA) and Context agent (CA). The agency is depicted in figure 5.

**Node Manager Agent (NMA):** This agent resides in all the sensor nodes of MSN. It creates context agent and NBB and is responsible for synchronizing the actions of the agents within themselves and outside world/agents. CA senses the image and updates the NBB periodically with image, capture time and image signal strength. NMA compares the sensed image with previous image and set of critical images residing in NBB and interprets the context. In sleep mode, NMA does not transmit any information to sink. The agent also monitors the battery life; if battery is exhausting, sends the status of the battery to its sink node. NMA sends information of the node such as node id, geographical information, context, and signal strength information to the sink node, if and only if it has a signal strength above the predefined threshold set by the sink (varies with time and critical situations).

**Node black Board (NBB):** This knowledge base is read and updated by the agents. NBB comprises of node id, active mode /sleep mode, sensing time, previously sensed image, set of critical images, bandwidth required to transmit in the available bandwidth, image signal strength and geographical status, and location of the node. Critical images comprises of images of weapons, explosives, enemy dresses, etc. Figure 6 depicts the sample NBB at 11.13AM, when the CA senses the image and updates the NBB. The object is detected to be a context-general object (Cgo).

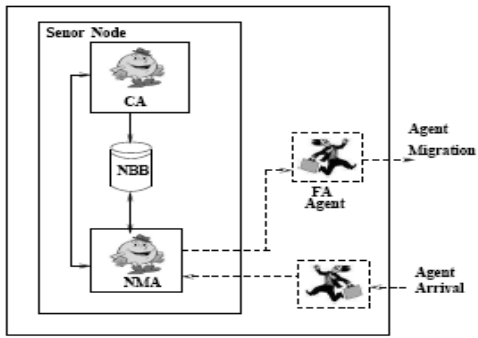

| Node ID | Location | Status | Battery in volts | Signal Strength | Power in millivolts | Critical Images | Previous Image | BW Required to Tx | Context |
|---|---|---|---|---|---|---|---|---|---|
| 3 | (6,13) | Active | 8.3 | 79% | 14.2 | cri1.tif cri2.tif cri3.tif cri4.tif cri5.tif cri6.tif cri7.tif | p1.tif | 13% | 11.13AM Cgo |

Figure 5: Node Agency                    Figure 6: Node Black Board

**Context Agent (CA):** It gathers the context from the target such as time of sensing the image and updates the NBB periodically. It interacts with NMA regarding the updating of the NBB. The agent is in direct contact with the sensor. It can be also triggered as and when required (aperiodic) based on the need of information by the sink.

## 1.5.2 Sink Agency

The sink agency comprises of static, mobile agents and sink black board (SBB). Agents employed are Fusing agent (FA), and Sink Manager Agent (SMA) which are mobile and static, respectively. Sink agency is shown in figure 7.

**Sink Manager Agent (SMA):** This agent resides in the sink node of MSN. It creates fusing agent and SBB and is responsible for synchronizing the actions of the agents within themselves and outside world/agents and monitors and updates the SBB continuously. It is responsible for the image fusion process. SMA gets fused images from the target in two ways: 1) based on the sensor node context (context driven) and 2) whenever user seeks (sink driven) fused information. SMA retrieves active sensor node's geographical locations information from the SBB, generates the optimum routes using standard WSN geographical routing protocol as and when required. Based on the context or as and when user needs the information, SMA triggers FA with the necessary fusion code, forward and reverse route. The fused image is given to the user monitoring the network either by alert alarm or updated in the user database of images.

**Fusing Agent (FA):** It is a mobile agent equipped with image fusion code (low/high resolution based on the context) that migrates from one to another active node depending on the routing information provided by the SMA. Whenever it visits active nodes, fuses the image, and moves to another active node along with the fused image. The agent repeats fusion until it visits all active nodes. After completion of the fusion process, FA sends the fused image to the sink node using reverse route specified by SMA.

**Sink Black Board (SBB):** It is the knowledge base that can read and updated by SMA. It stores the information about the node id, signal strength, context information, time of sensing, image signal strength, bandwidth required to transmit the image of each active node, available network bandwidth, and geographical locations of the active nodes. Sample contents of SBB is shown in the figure 8 (using 1 as the sink node).

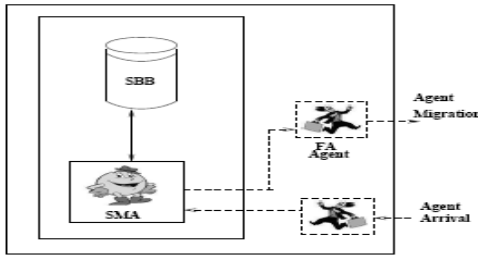

| Node ID | Location | Status | Signal Strength | Battery in voln V | Power in milliwatts | BW Required to Tx | Context |
|---|---|---|---|---|---|---|---|
| 3 | (x,y) | Active | 7900 | 8.5 | 3.1 | 1500 | 11.15AM Cgo |
| 6 | | Inactive | 000 | | | | |
| 10 | | | | | | | |
| 14 | | | | | | | |
| 17 | | | | | | | |
| . | | | | | | | |
| . | | | | | | | |
| . | | | | | | | |

Figure 7: Sink Agency                    Figure 8: Sink Black Board

## 1.6 Agency Algorithm

In this section we discuss the node and sink agency algorithms.

### 1.6.1 Node Agency Algorithm

Nomenclature:x,y: index value of the image, $D(x; y)$ = Difference image,$C(x; y)$= Critical image, $P(x; y)$= Present image, $B(x; y)$= Previous image, $C_{go}$= General object context, C$co$= Critical object context, $i$: index value of the critical images, , $N\_N$: Size of the image,and $H$: Entropy operator, $H1$;$H2$=Entropy of present and previous images, $Tth$= Threshold of image signal strength, $Pth$ =Percentage of similarity between previous and present image, and $Nstatus$= Node status, N$criim$= Number of critical images.

**Begin**

    1. CA gathers the image and time of sensing
    2. NMA gets the context information and does the interpretation

        For x= 1 to N do,
        For y= 1 to N do,
        H1=H(P(x,y));
        H2=H(B(x,y));
    endfor
    endfor

    $P_{th} = (H1/H2)$X100;

    NMA updates the $Pth$ to NBB;
    if($Pth > Tth$)

        $N_{status}s = Active$;
        NMA updates the node status to NBB;
  1. else

        $N_{status} = Inactive$;
        Delete the image from NBB;
        NMA updates the node status to NBB;
        Go to (4);
        For i= 1 to N$criim$,
        For x= 1 to N do,
        For y= 1 to N do,
            D(x,y)= $P_{im}$(x,y) - $B_{im}$(x,y);
        if(D$i$(x,y) = =C$i$(x,y)) Then,
            Context= C$co$;
        else
            context =C$go$;
        endfor
    endfor
    endfor

    3. NMA updates the context information to NBB;
    4. Stop

**End**

CA gathers the context such as image and sensing time and updates the NBB. NMA finds the similarity between the previous and present image by taking the entropy ratio of them. The percentage of similarity ratio compared with $T_{th}$, if it is greater than the threshold value then node is assigned with active status otherwise inactive status is assigned. If the node is inactive then present image is deleted from NBB. The D(x,y) is compared with set of critical images, C(x,y) of the node and if any critical object is found, then it is assigned with $C_{co}$, else $C_{go}$. NMA also finds the entropy of the difference image and compares with threshold signal strength of an image, if it is greater; than status of the node is stored as active else status of the node is stored as inactive. NMA updates the context information to NBB and sends it to the sink node by using the flooding mechanism. In flooding mechanism message is sent from node to node until it reaches the sink node.

## 4.6.2 Sink Node Agency Algorithm

Nomenclature:F(x,y) = Fused Image, N(x,y)= Active node Image, $N_{active}$= Number of active nodes $C_{go}$= General Object Context, $C_{co}$= Critical Object Context, $R_{table}$= Routing table to reach active sensor node, i=

index value of the active nodes, x,y= index value of the node image, $F_{gocode}$= General objection detection image fusion code, $F_{cocode}$= Critical objection detection image fusion code $NXN$= Size of the image

.

**Begin**

1. SMA gets the context information from SBB;

2. SMA creates FA, and provides the context information;

3. Based on the Context FA gets the fusion code and routing table

        if (context= = C$go$) Then,

                FA gets the $F_{gocode}$ and $R_{table}$;

        else

                FA gets the $F_{cocode}$ and $R_{table}$;

4. Along with fusion code and routing information, FA visits active sensor nodes for fusion

        F(x,y)=0;

        For i=1 to $N_{active}$

        For x= 1 to N do,

        For y= 1 to N do,

        F(x,y)= F(x,y) + N$i$(x,y);

        endfor

        endfor

        endfor

5. FA sends the fused image to the sink node;

6. FA is disposed;

7. Go to (1) as and when the user seeks the information and repeat the process;

8. stop;

**End**

Whenever the SMA gets the context information or user seeks information, it creates the FA, based on the context; FA retrieves low/ high resolution fusion code. With fusion code FA visits all the active sensor nodes to fuse the images, and returns with fused image to the sink node. Fusion of images is done using DWT, which is discussed in the section 3.2. For simplicity, in the sink agency algorithm fusion has considered to be an addition operation.

## 4.7 Agent Interaction

Context based image fusion system agent interaction sequence is depicted in figure 9, which provides a detailed view of the image fusion by agents in a MSNs. The numbers shown on the directed arcs denotes the action number in the sequence of interactions that takes place.

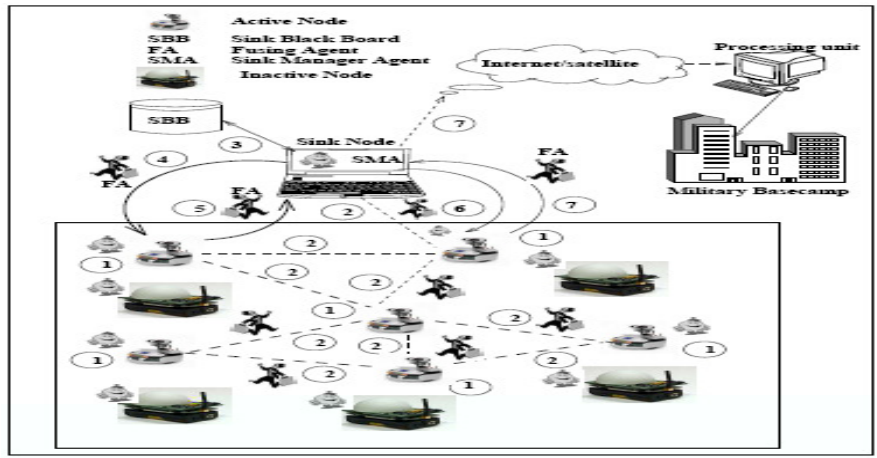

Figure 9: Information Fusion System Agent Interaction

The interaction sequence is as follows:

1.  CA of the active sensor nodes gathers the context namely sensing time and image. NMA compares the sensed image with the previous and set of critical images of the node and interprets the context.
2.  NMA sends the context information into the sink by using the flooding protocol.
3.  NMA updates the context information to SBB.
4.  SMA gets the context information, creates the FA, along with fusion code (low/high resolution) FA visits the first active node and fuses the image and migrate to the next active node and continue the process till it visits all the active nodes.
5.  FA returns to sink node along with fused image.
6.  During night time, as and when user requires fused information, SMA generates the FA agent and sends it for the fusion.
7.  FA returns with fused image to sink node. Fused image will be sent to the military base camp through the Internet or satellite.

## 5. Simulation model and performance metrics

We have carried out the simulation of the *context aware image fusion using multi-agents for MSNs*. The proposed model has been simulated in various network scenarios on a Pentium-4 machine by using MATLAB tool for the performance and effectiveness of the approach.

**Simulation Model**

Proposed scheme is simulated in various network scenarios to access the performance and effective of an approach. Simulation environment comprises of six models namely network model, Channel Model, Propagation model, Battery model, CAC model, and information fusion model. These models are described as follows.

**1. Network Model:**

We considered simulation area of $A$ X B sq meters with LOC scenario. A MSN consists of *num* static nodes that are placed randomly within the given area. MSN have communication radius, $r$ mts, and network bandwidth, *net_BW*. $F_{code}$ kbytes wavelet based image fusion code roams around the network.

## 2. Channel Model

The communication environment is assumed to be contention-free (e.g., a media access scheme such as time division media access (TDMA) may be assumed). The transmission of packets is assumed to occur in discrete time. A node receives all packets heading to it during receiving interval unless the sender node is in off state. For simplicity, we have considered the channel to be error free. The characteristics of sensor networks and applications motivate a MAC that is different from traditional wireless MACs such as IEEE 802.11 in almost every way: energy conservation and self-configuration are primary goals, while per-node fairness and latency are less important. Sensor MAC protocol (S-MAC) uses different techniques to reduce energy consumption and support self-configuration. To reduce energy consumption in listening to an idle channel, nodes periodically sleep. Neighboring nodes form virtual clusters to auto-synchronize on sleep schedules. Simulation environment uses the S-MAC protocol.

## 3. Propagation model

Free space propagation model is used with propagation constant β. Transmission range of WMSN node communication radius is *r* for a single-hop distance *d*. It is assumed that at any given time, the value of transmitted power is $N_{Pow}$ *milliwatts* for every node.

## 4. Battery model

In MSN, image sensor nodes are deployed in the battle field, recharging of the nodes at the target is difficult. So, we have used a solar cell recharging model [36] and a layered clustering model to deal with the restrict energy consumption under the consideration of visual quality. The system lifetime can be prolonged by rechargeable solar cell that can be recharged by solar panel in daytime. Image sensor nodes consumes *node-batt millivolts* (day, non critical information= less, day critical information= medium, and night=more) to sense an image.

## 5. CAC Model

Various contexts considered are: General object detection context, Critical object detection context and night detection context. The images and their respective contexts are stored in the knowledge base of each sensor. Contexts are randomly generated by assigning a number such that 1= General object detection, 2 = Critical object detection, 3 = Time (night or poor lighting condition) context.

## 6. Information Fusion Model

We are using two information fusion methods namely context driven image fusion and user driven image fusion. Context driven fusion uses low resolution image fusion or high resolution image fusion. User/sink driven image fusion uses the high resolution image fusion. Each of the sensor node is associated with battery of *node-batt millivolts*. It is assumed that 1 millivolt is decremented for every usage (transmission/processing). *k* number of active nodes are randomly chosen as active nodes from *num* nodes. Set of critical images, present and previous images are stored in the node. Size of the gray images stored at each of the sensor node is fixed. The fixed gray scale image of size *rows X columns, 8 bits/pixel* is assigned to each of the active sensor node. *Th*, is the percentage of the threshold image signal strength.

## 5.1 Simulation Procedure

To illustrate some results of the simulation, we have taken *A* = 100, *B* =200 sq meters and N=1 to 5, *num*=5 to 15, *netBW*=40mbps, *node_batt*=90 millivolts, Th= (50%, 60 %, 70 %), Gray scale image of varying size *rows_columns*=(32X32; 64X64; 128X128; 256X256), (8,12,16,24) bits/pixel, time of sensing (*Tsensing*= (8*AM*, 11.30*AM*,5*PM*, 7*PM*), *Pth*= Present Signal strength (30%,50%,60%,70%), communication radius *r*=10 *mts* , *netBW*= 4*MBPS*, Propagation Constant β =3.5, and *Fcode* = (4 , 8, 12) *Kbytes*. Each of the simulation executes in varying seconds.

**Begin**

- *Generate the WMSN for the given radius and number nodes.*
- *Apply the proposed context aware fusion model.*
- *Compute the performance of the system.*

**End**

Performance parameters considered in the simulation are as follows.

1. **Node Battery Usage:** It is defined as the battery depleted with the usage of a node.
2. **Dropping rate:** It is defined as the number of packets missed during the transmission of packets (Equation 9), i.e., it is a ratio of packets dropped to packets sent.

$$DroppingRate = \frac{PacketSent - PacketReceived}{TotalPacketssent} \tag{9}$$

1. **Fusion time:** It is the total time required by the FA to fuse the images from active sensor nodes.
2. **Bandwidth Requirement:** It is the amount of bandwidth required to transmit the image to sink node, i.e., it is a ratio of image size to available bandwidth. It is given by (Equation 10)

$$Brequired = \frac{Sizeoftheimage}{AvailableBandwidth} \tag{10}$$

3. **Throughput:** It is the ratio of Image packets (data) received to the image packets sent and is given by (Equation 11).

$$Throughput = \frac{Imagepacketsreceived}{Imagepacketssent} \tag{11}$$

4. **Mean Square Error:** It is defined as the standard deviation of the difference image between the ideal and standard image.
5. **Agent Overhead:** It is the additional code which acquires the communication channel. It is given by (Equation 12).

$$AgentOverhead = \frac{Imagesize}{SizeofAgent + Imagesize} \tag{12}$$

## 6. Results

In this section, we discuss the various results obtained through the simulations. The results include node battery usage, dropping rate, throughput, fusion time, bandwidth requirement, agent overhead and mean square error.

## 6.1 Node Battery Usage

The life time of the sensor node mainly depends on the battery life time and its power. Sensor nodes will be active whenever the sensor nodes have the information otherwise nodes status will be inactive (sleep mode). In active mode of the sensor node, sensor nodes consume more power and in inactive mode, a node consumes less power. This is evaluated by choosing the one sensor node of WSN for repeated simulation. For each of the simulation, sensor node sends the varying number of packets. We observe from the figure 10 that the battery life decreases as the number of packets sent by the sensor node increases.

Sensor nodes require more power to transmit the critical and dark images which are of large size (better resolution) as compared to day time image. During day time, only low resolution images are to be sent to the sink node that needs less power. From figure 11, we can notice that power consumed by each of the sensor node to transmit the day time, non critical image is less as compared to the critical dark images.

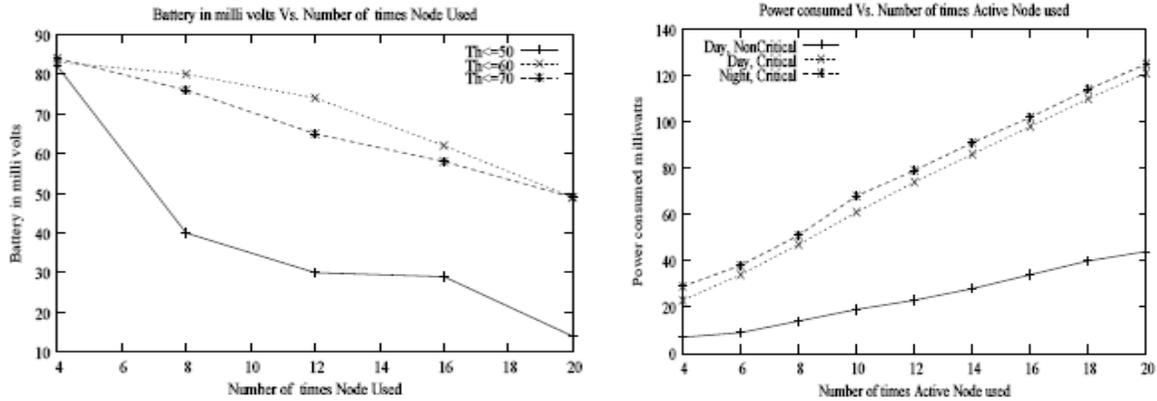

Figure 10: Battery in Milli Volts Vs. Number of times node used, Figure 11: Power in Milli watts Vs. Number of times node used

## 6.2 Dropping Rate

During the transmission of the image packets in the network, some packets may not reach the sink node. Dropping rate depends on the number of packets dropped. From figure 12, we can notice that dropping rate increases as there is increase in the number of nodes and decrease in signal strength value (Th).

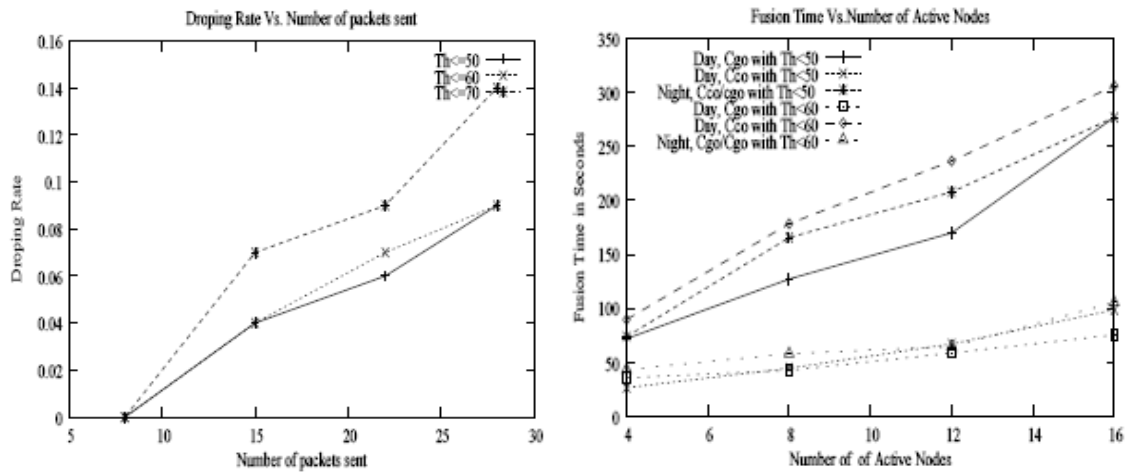

Figure 12: Dropping Rate Vs No. of Active Nodes, Figure 13: Fusion Time in milliseconds Vs. No. of Active Nodes

## 6.3 Fusion Time

Since the sensor nodes have less processing capability, mobile agent requires more time to fuse the image from all the active nodes.

If there is day time and non critical images in all the active nodes, then all the active nodes information has to be fused and sent to the sink node which needs less time as we use low resolution image fusion technique and better lightning condition. In case of critical object detection context, if the lighting condition is poor, we use high resolution image fusion method which needs more time for fusion. From the figure 13, we depict that for critical images as the number of nodes and *Th* increases, the fusion time also increases. We can notice that fusion time required for the noncritical object detection context is less compared to the noncritical object detection context.

## 6.4 Bandwidth Requirements

We allocate 8 or 12 bits/pixel for the low resolution image. For high resolution image, we allocate 15 or 24 bits/pixel depending on the availability of the bandwidth. For critical and dark images, we allocate more number of bits compared to the general object detection context images.

## 6.5 Throughput

Throughput depends on number of image packets sent and received. In this paper, we have considered the gray scale image. While transmitting the image, it is divided into packets and sent. The figure 16 shows that it decreases with increase in the image size

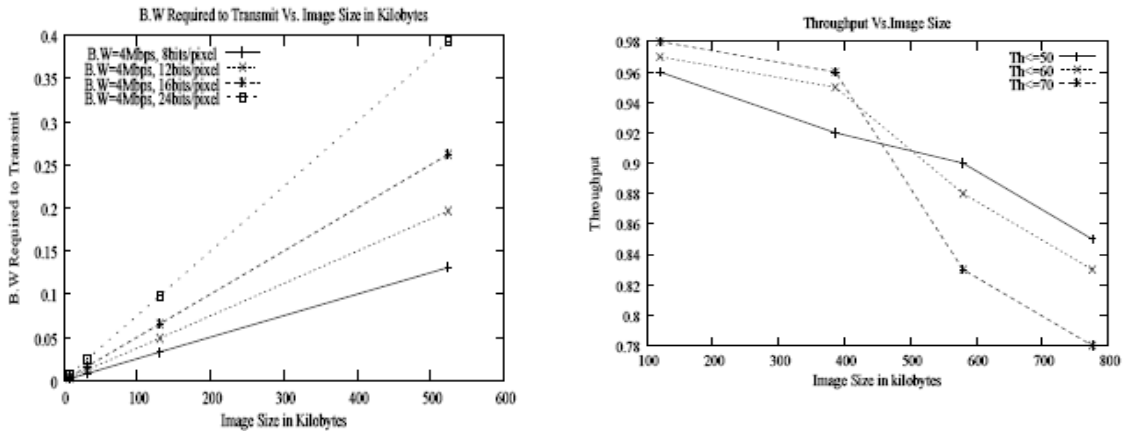

Figure 15: BW  Vs. Size of the image,  Figure 16: Throghput Vs. Image size.

## 6.6 Agent Overhead

Agent overhead depends on the size of agent. Fusing agent has to carry image along with the wavelet based image fusion code. Agent code is extra overhead to the communication channel. Figure 17 shows that as the image size increases, the overhead decreases, as the same size agent is used for handling small as well as large size image. We can also observe, as the size of the agent increases agent overhead also increases.

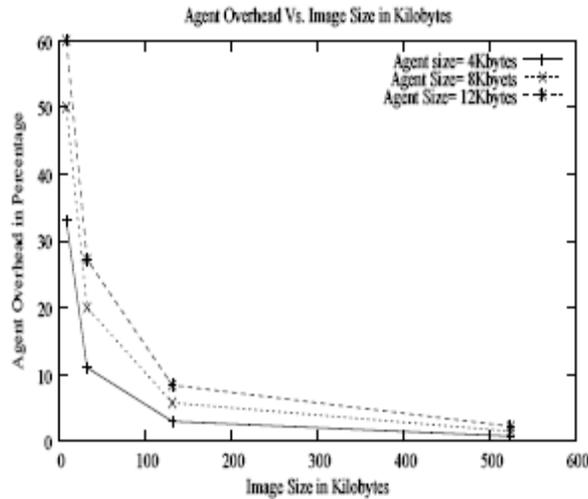

| Wavelet Bases | MSE |
|---|---|
| db3 | 50 |
| db4 | 52 |
| db10 | 48 |
| Bior1.1 | 60 |
| Bior1.3 | 51 |
| Bior1.5 | 56 |
| Bior2.4 | 48 |
| Bior3.7 | 47 |
| Bior4.4 | 51 |
| Haar | 60 |
| Mayer | 47 |

Figure 17: Agent Overhead Vs. Image Size, Table 1: MSE for different Wavelet Bases

## 6.7 Mean Square Error

We have also analyzed the performance of the image fusion in terms of mean square error (MSE) ρ for different wavelet bases such as biorthogonal, db, haar, and Mayer. ρ is defined as the standard deviation of the difference image between the ideal and standard image  (See  table 1).

# 7. Conclusion

This paper proposed a context aware agent based distributed sensor network (CADSN) to form an improved infrastructure for multi-sensor image fusion to monitor the militant activities. The proposed work is based on context aware computing which uses software agents for image fusion in WMSN. In an environment where source nodes are close to each other, and considerable redundancy exists in the sensed data, the source nodes generate a large amount of traffic on the wireless channel, which not only wastes the scarce wireless bandwidth, but also consumes a lot of battery energy. Instead of each source node sending sensed images to the sink node, images from the different active nodes are fused and sent to sink node by using mobile agent. The use of agents facilitates the following: 1) asynchronous operation, i.e., does not require a continuous connectivity between source and sink, 2) flexibility to change the embedded code to perform context/user driven fusion, 3) adaptability to varying network conditions and the environment for image fusion, 4) ease of maintenance since the code can be debugged and upgraded independent of other agents in the system, 5) reusing of the code is possible by other applications with slight modifications and put in the system, thus enables dynamic software architecture.

However, there are some issues to be questioned in the proposed scheme, which can be taken up as further work: security in information fusion by mobile agent, standard agent framework supporting persistence and security to agents, tackling the active node failures during agent's fusion process?, positioning   of the cameras concerned to images to be fused, and security to the sensors. Furthermore, we are planning to employ cognitive agents to tackle the aforesaid questions.

## ACKNOWLEDGMENTS

We are thankful to Visvesvaraya Technological University (VTU), Karnataka, INDIA for sponsoring the part of the project under VTU Research Grant Scheme, grant no. VTU/Aca/2009-10/A-9/11624, Dated: January 4, 2009.

## Authors

**A. V. SUTAGUNDAR** completed his M. Tech from Visvesvaraya Technological University, Belgaum, Karnataka. He is pursing his PhD in the area of Cognitive Agent based Information Management in wireless Networks. Presently he is serving as an Assistant Professor of Department of Electronics and Communication Engineering Bagalkot, Karnataka. His areas of interest include Signal and system, Digital Signal Processing, Digital Image Processing, Multimedia Networks, Computer communication networks, Wireless networks, Mobile ad-hoc networks, Agent technology. He has published 16 papers in referred National/International Conferences and 4 papers in international journals.

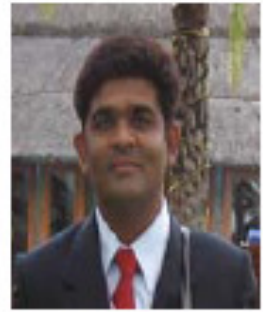

**Sunilkumar S. Manvi** received M.E. degree in Electronics from the University of Visweshwariah College of Engineering, Bangalore and Ph.D degree in Electrical Communication Engineering, Indian Institute of Science, Bangalore, India. He is currently working as a Professor and Head of Department of Electronics and Communication Engineering, Reva Institute of Technology and Management, Bangalore, India. He is involved in research of Agent based applications in Multimedia Communications, Grid computing, Vehicular Ad-hoc networks, E-commerce and Mobile computing. He has published about 100 papers in national and international conferences and 40 papers in national and international journals. He has published 3 books. He is a Fellow IETE (FIETE, India), Fellow IE (FIE, India) and member ISTE (MISTE, India ), member of IEEE (MIEEE, USA),
He has been listed in Marqui's Whos Who in the World.

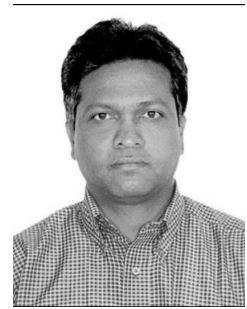